\documentclass[final,5p,times,twocolumn]{elsarticle}

\usepackage{hyperref}
\usepackage[dvipsnames]{xcolor}
\usepackage{algpseudocode}
\usepackage{algorithm}
\usepackage{graphicx}
\usepackage{adjustbox}
\usepackage{amsmath,amssymb} 
\usepackage{tabularx}
\usepackage{multirow}
\usepackage{subcaption}
\usepackage{soul,color}
\usepackage{float}
\usepackage{graphicx,wrapfig,picins}
\usepackage{caption}
\usepackage{enumitem}
\usepackage{array}
\usepackage{booktabs}
\usepackage{dblfloatfix}
\usepackage{float}
\usepackage{bbding}
\usepackage{rotating}
\usepackage{tablefootnote}

\begin{document}

\begin{frontmatter}

\title{A Secure and Interoperable Architecture for Electronic Health Record Access Control and Sharing}

\author[mythirdaddress]{Tayeb Kenaza}\corref{mycorrespondingauthor}
\cortext[mycorrespondingauthor]{Corresponding author}
\author[mythirdaddress]{Islam Debicha}
\author[mythirdaddress]{Youcef Fares}
\author[mythirdaddress]{Mehdi Sehaki}
\author[mythirdaddress]{Sami Messai}


\journal{Elsevier Journal}

\begin{abstract}

Electronic Health Records (EHRs) store sensitive patient information, necessitating stringent access control and sharing mechanisms to uphold data security and comply with privacy regulations such as the General Data Protection Regulation (GDPR). In this paper, we propose a comprehensive architecture with a suite of efficient protocols that leverage the synergistic capabilities of the Blockchain and Interplanetary File System (IPFS) technologies to enable secure access control and sharing of EHRs. Our approach is based on a private blockchain, wherein smart contracts are deployed to enforce control exclusively by patients. By granting patients exclusive control over their EHRs, our solution ensures compliance with personal data protection laws and empowers individuals to manage their health information autonomously. Notably, our proposed architecture seamlessly integrates with existing health provider information systems, facilitating interoperability and effectively addressing security and data heterogeneity challenges. To demonstrate the effectiveness of our approach, we developed a prototype based on a private implementation of the Hyperledger platform, enabling the simulation of diverse scenarios involving access control and health data sharing among healthcare practitioners. Our experimental results demonstrate the scalability of our solution, thereby substantiating its efficacy and robustness in real-world healthcare settings. 
\end{abstract}


\end{frontmatter}


\section{Introduction}
\label{intro}
Electronic Health Records (EHRs) play a crucial role in modern healthcare services, serving as a standardized information model that integrates the patient's health information over time. The use of information systems for managing patient data within healthcare providers has proven to be beneficial in terms of time and cost reduction. However, the increasing demand for remote access to EHRs from different health providers, as well as from patients' homes, poses new challenges regarding security and interoperability between healthcare actors \cite{blobel2018interoperable}.

Several solutions have been proposed to secure the sharing of EHRs between actors, with cryptography and access control being the primary approaches \cite{rezaeibagha2016distributed,anderson2017information,ying2018lightweight,qiu2020secure,dubovitskaya2017secure,ming2018efficient}. However, traditional cryptographic solutions, such as public-key cryptography, are complex to implement within distributed sharing systems and often require a public key infrastructure (PKI) for effective key management \cite{qiu2020secure,dubovitskaya2017secure}. This latter, while effective in many contexts, often falls short in providing the desired level of patient-centric control and ensuring compliance with evolving privacy regulations like the General Data Protection Regulation (GDPR). These limitations have prompted the need for alternative approaches that can address the main requirements of healthcare applications while ensuring secure and controlled access to EHRs.

In this context, blockchain technology has emerged as a promising solution for addressing the challenges associated with EHR access control and sharing \cite{shahnaz2019using,mayer2020electronic}.  We deliberately emphasize blockchain technology as a foundational component for enhancing  EHR access control and sharing because of its unique features that align with the specific requirements of healthcare data management. Indeed, blockchain's decentralized and immutable nature ensures data integrity and tamper-proof audit trails, crucial for maintaining the trustworthiness of health information.

In this paper, we propose an architecture based on a private blockchain and InterPlanetary File System (IPFS) as a reliable and robust solution for a secure and flexible sharing of health records. This approach allows patients exclusive control over their EHRs through smart contracts. This decentralized model aligns better with the vision of empowering individuals to manage their health information autonomously, a feature not readily achievable with traditional PKI-based systems. Moreover, the immutability and transparency of blockchain transactions enhance the auditability and accountability of EHR access, addressing the specific security and privacy needs inherent in healthcare data management.  

However, integrating blockchain technology into an already operational healthcare information system is not a simple task. Indeed, enhancing agility and efficiency, reducing the number of requests to the blockchain, reducing the amount of stored data in the blockchain, and securing off-chain communication are the key challenging features that we considered in our proposal.

Our proposed architecture leverages the functionalities offered by blockchain and IPFS technologies to facilitate secure access and sharing of health data while adopting decentralized data storage within healthcare providers. The architecture comprises a set of interoperable protocols, each assigned with specific tasks and responsibilities. These protocols encompass key functions expected from an EHR sharing system, such as actor registration, data integrity verification, access control, and secure EHR consultation.

The deployment and operation of private blockchains in our proposed architecture involve the creation of a permissioned network, accessible only to authorized participants, typically healthcare providers and patients. This approach ensures a controlled environment, aligning with the sensitivity of health data. Smart contracts are deployed on this private blockchain to enforce access control exclusively by patients, fostering compliance with GDPR and other data protection regulations. The utility of blockchain technology in our context is rooted in its ability to provide a transparent, decentralized, and tamper-proof ledger, which is crucial for maintaining the integrity of EHRs. The immutability of blockchain transactions ensures an auditable trail, contributing to the secure and accountable sharing of health information. Here, the use of IPFS is instrumental in addressing data heterogeneity challenges and ensuring the secure and decentralized storage of EHRs outside of the blockchain. This is crucial because a huge EHR cannot be transmitted over the blockchain.

Additionally, we present in this paper a demonstrative prototype that showcases the operational functions of our proposed architecture. The prototype simulates various use cases involving actors in the healthcare sector, illustrating the implementation of different aspects of the architecture, including user applications, smart contracts for interfacing with the blockchain, and communication layers.

The remainder of this paper is organized as follows. Section 2 gives a background on the challenges of healthcare data sharing and briefly discusses some related works. Section 3 highlights the potential of blockchain and IPFS in healthcare data management. Section 4 presents an overview of our architecture for secure EHR sharing and access control. Section 5 covers the main protocols used by our architecture, and Section 6 shares our experiment results. Section 7 discusses open issues, aiming for a comprehensive understanding. Finally, Section 8 concludes the paper.

\section{Background}
\label{sec:2}
In the rapidly evolving landscape of healthcare, the efficient and secure sharing of EHRs is paramount for delivering optimal patient care. However, data sharing in healthcare poses significant security challenges, including data breaches and unauthorized access. An exploration of related works seeks to provide a robust framework to revolutionize healthcare data management, promoting seamless interoperability and safeguarding sensitive patient information.

\subsection{Data sharing in healthcare}
\label{sec:2.1}

Patient records serve as comprehensive repositories of administrative and medical information of individuals, with the transition from paper-based to digital records driven by advancements in computer technologies and networks. Electronic medical records serve as digital counterparts to paper charts in clinicians' offices, containing medical and treatment histories specific to a particular disease. However, electronic medical records face limitations in terms of easy data mobility outside the clinical setting, sometimes necessitating physical delivery to specialists and other care team members. In contrast, EHR systems aim to surpass the confines of standard clinical data captured in provider offices, enabling the seamless sharing of medical information among stakeholders. The advantages of EHRs include reducing redundancy efforts, improving efficiency in data gathering and accessibility, tracking information usage, minimizing professional errors, and generating a vast corpus of medical data for scientific research and overall enhancement of patient care~\cite{venot2013informatique}.

Efficient sharing of EHRs is crucial for coordinating available resources to meet the increasing healthcare needs of the population. However, this necessitates the creation, development, and higher-level management of EHRs beyond individual health providers, enabling simple and efficient utilization within specific regions or territories. One of the primary challenges in EHR sharing is the heterogeneous nature of health data formats employed by various actors in the medical field. The diverse representation of data for the same document type hampers seamless sharing among different entities. To address this challenge, standardization efforts are underway by institutions such as HL7 (Health Level Seven International), CEN/TC 251 (European Committee for Standardization), ISO-TC-215 (International Organization for Standardization, Health Informatics), and DICOM (Digital Imaging and Communications in Medicine) to promote harmonized health data exchange~\cite{conley2018gdpr}.

These international institutions are actively advancing numerous aspects of electronic healthcare, with a particular focus on sharing EHRs to achieve personal, professional, social, and economic objectives. Another significant challenge lies in ensuring secure and beneficial data sharing for professionals and practitioners, facilitating seamless coordination of healthcare by providing access to all relevant information, thereby enhancing decision-making processes.

\subsection{Security challenges}
\label{sec:2.2}
The rapid development and computerization of EHR have presented a dilemma for the global healthcare system. On one hand, this advancement offers facilitation and optimization, while on the other hand, it introduces security risks associated with the integration of IT tools in the healthcare sector. Given the substantial benefits of EHRs, it is imperative to identify and understand the security challenges \cite{chenthara2019security,christiansen2017shared} arising from the emergence of new technologies in the medical domain and propose effective solutions to address them.

One significant challenge in shared EHR systems is the limitation of roles and responsibilities. Unlike traditional healthcare systems with well-defined responsibilities based on physical boundaries, shared healthcare systems raise legal and ethical concerns due to the ambiguity in role delineation. Another major difficulty lies in the lack of integration between global and local systems. This integration gap is often caused by inadequate internal system documentation or the absence of a supervised and secure access control system.

To contribute to addressing these challenges, our focus is on developing secure and efficient protocols for access control and the sharing of EHRs. Instead of relying on traditional approaches that involve a trusted third party, we have chosen to leverage blockchain technology. By utilizing blockchain, we aim to achieve decentralized, easily integrable, and secure management of health data. This approach offers the advantages of increased transparency, immutability, and robustness in data handling, which are essential for ensuring the privacy, integrity, and availability of EHRs. In addition, to overcome the blockchain storage limitation, we use the IPFS technology to store EHRS outside the blockchain.

\subsection{Related works}
\label{sec:2.3}

In this section, we present a short analysis of recent works in blockchain-based Electronic Health Record (EHR) systems. Following the review of the 22 selected papers, we were able to affirm that blockchain-based EHR systems exhibit a wide range in architectural and design choices. To capture this diversity and provide a structured foundation for comparative analysis, we synthesized recurring features and patterns at the conceptual level. This process led us to identify eleven distinct properties that consistently appear within the literature in the proposed models and that tackle the core functionalities of BC-EHR systems.

The comparative Table \ref{tab:taxonomy-comparison} highlights the architectural diversity and evolving priorities in blockchain-based EHR systems. By systematically analyzing 22 contributions across eleven dimensions—including storage architecture, blockchain type, consensus, access control, privacy techniques, standards compliance, and consent models—we extract key insights and expose unresolved challenges that inform the broader trajectory of BC-EHR research.

A clear trend across the reviewed systems is the widespread adoption of hybrid storage architectures, particularly two-layered approaches that combine on-chain metadata management with off-chain data storage in IPFS, cloud services, or hospital databases. This reflects a pragmatic balance between decentralization and scalability, driven by the inherent limitations of storing large medical files on-chain. However, only a minority of systems employ three-layered or multi-chain models—such as those involving BigchainDB or zk-Rollups—despite their potential to enable more granular control, dynamic policy enforcement, or immutable high-throughput logs. This suggests a gap in the exploration of modular, performance-scalable storage stacks tailored for complex medical workflows.

Regarding blockchain type, the table confirms a strong shift toward permissioned networks, typically consortium or private blockchains. This aligns with healthcare-specific needs such as auditable governance, fine-grained access control, and institutional accountability. Hyperledger Fabric emerges as a preferred platform, though Ethereum remains in use for its programmable smart contract layer. Notably, interoperability across platforms (e.g., Ethereum + IPFS or Fabric + BigchainDB) is rare, and few systems explore bridging mechanisms or middleware to unify different ledgers, pointing to a gap in cross-chain EHR data orchestration.

In the consensus dimension, most permissioned systems employ RAFT or Kafka for fault-tolerant, low-latency consensus, while public systems rely on PoW or PoA. However, few papers empirically compare consensus impacts on system performance, especially in healthcare-specific conditions (e.g., emergency overrides, delayed replication, or audit throughput). This lack of performance-driven evaluation constitutes a methodological gap and calls for more benchmarking of consensus under realistic clinical load.

The access control dimension reveals considerable heterogeneity. While RBAC remains prevalent due to its simplicity, it is often insufficient for representing the complexity of healthcare access scenarios. ABAC models, capable of policy enforcement based on roles, time, department, or user attributes, are increasingly adopted, particularly in Fabric-based systems. PBAC and IDAC appear in fewer but more advanced architectures, enabling policy-driven or self-sovereign control over data access. However, most systems implement access control at the smart contract or chaincode level, and very few link policy enforcement to patient-defined rules or external consent authorities. Furthermore, despite the potential of decentralized identity (e.g., DIDs, VCs), integration of identity frameworks remains limited, marking an important direction for future development in achieving decentralized, verifiable, and revocable access control.

In the realm of privacy and security, the comparative table shows that cryptographic primitives like ZKPs, ABE, and pseudonymization are beginning to appear, albeit sparsely. Though ZKPs are used in only three systems and ABE in two, this indicates early experimentation rather than widespread adoption. Proxy re-encryption and homomorphic encryption—while theoretically attractive—are not yet implemented in any of the reviewed systems. TEEs are used in just one case. Despite the active research into cryptographic techniques observed during the screening phase, the sparse implementation of advanced primitives in the reviewed systems reveals a significant innovation gap in deployable privacy-enhancing technologies, particularly for scenarios demanding real-time consent, computation on encrypted data, or policy-proof access without confidentiality breaches.

\begin{sidewaystable*}[htbp]
\caption{Comparison of the reviewed papers}
\label{tab:taxonomy-comparison}
\centering
\resizebox{\linewidth}{!}{%
\renewcommand{\arraystretch}{1.7}
\begin{tabular}{|p{1 cm}|p{3.5cm}|p{2.2cm}|p{2.5cm}|p{2.2cm}|p{3.5cm}|p{3.8cm}|p{2cm}|p{2.5cm}|p{2.8cm}|p{2.8cm}|p{2cm}|}
\hline
\textbf{Paper} & \textbf{Storage Model} & \textbf{Blockchain Type} & \textbf{Blockchain Platform} & \textbf{Consensus Algorithm} & \textbf{Access Control (Model \& Layer)} & \textbf{Privacy/Security Techniques} & \textbf{Maturity Level} & \textbf{Standard Conformance} & \textbf{Regulation Compliance} & \textbf{Stakeholders} & \textbf{Consent Model} \\
\hline
\cite{PbDinEHR} & Three-layered (Ethereum + SQL + IPFS) & Permissionless-Public & Ethereum & PoA & PBAC (SC-level) & Hybrid Encryption, PbD, Data Masking & Prototype & ISO/IEC 27001, 29100, 15288 & GDPR, APPs & Patients, Doctors, Institutions, Gov. & Object-Oriented \\
\hline
\cite{ACTION-EHR}& Two-layered (HLF + IPFS) & Permissioned-Consortium & Hyperledger Fabric & RAFT & PBAC (MSP + CC) & Encrypted Off-Chain, FHIR, Audit Logs & Prototype & FHIR & GDPR, HIPAA & Patients, Doctors, Institutions & Dynamic \\
\hline
\cite{ZeroTrustBlock} & Three-layered (HLF + IPFS + BigchainDB) & Permissioned-Private & Hyperledger Fabric & RAFT & PBAC + ABAC (MSP + CC) & ZKP, ABE, TEE, Audit Trails & PoC & FHIR (partial) & HIPAA (partial) & Patients, Doctors, Insurers & Dynamic \\
\hline
\cite{Smitha2023} & Two-layered (HLF + IPFS) & Permissioned-Private & Hyperledger Fabric & RAFT & ABAC (MSP + CC) & Pseudonymization, CA/MSP & Prototype & Not Specified & Partial GDPR & Patients, Doctors, Hospitals & Dynamic \\
\hline
\cite{Sun2022} & Two-layered (HLF + IPFS) & Permissioned-Private & Hyperledger Fabric & Kafka & ABAC (CC) & ABE, Modular Chaincode & Prototype & Not Specified & Implied GDPR & Patients, Doctors & Dynamic \\
\hline
\cite{HealthChain} & Two-layered (Ethereum + Hospital DB) & Permissionless-Public & Ethereum & PoA & RBAC (SC) & Smart Contract, Encrypted DB & PoC & None & Not Specified & Patients, Doctors, Hospitals & Static \\
\hline
\cite{Nguyen2021} & Two-layered (Ethereum + Cloud) & Permissionless-Public & Ethereum & PoW & RBAC (Off-Chain) & Encrypted Storage, MEC Offload & Prototype & None & Partial GDPR & Patients, Doctors & Dynamic \\
\hline
\cite{Ma2024} & Three-layered (Ethereum + ZK-Rollup + IPFS) & Permissionless (L2) & Ethereum (zkRollup) & ZKP Proof-of-Validity & Conceptual & ZKP, Encrypted IPFS & Model & None & Not Specified & Patients, Researchers & Dynamic \\
\hline
\cite{Manoj2022} & Two-layered (Indy + Off-chain) & Permissioned-Private & Hyperledger Indy + Aries & n/a (VC Auth) & IDAC (VC + DID) & ZKP, VC, DID & PoC & None & GDPR (Implied) & Patients, Researchers & Dynamic \\
\hline
\cite{Barka2022} & Two-layered (Ethereum + Azure DB) & Permissionless-Public & Ethereum & PoW & RBAC + Biometric (SC) & Biometric Hash, AES & Prototype & None & Not Specified & Patients, Doctors & Emergency \\
\hline
\cite{Shuaib2022}& Two-layered (Ethereum + IPFS) & Permissionless-Public & Ethereum & PoW & RBAC (SC) & Digital Signatures & Prototype & None & Not Specified & Patients, Doctors & Static \\
\hline
\cite{Pilares2022} & Two-layered (Sawtooth + IPFS) & Permissioned-Consortium & Hyperledger Sawtooth & PBFT & RBAC (SC + Off-chain) & Audit Logs, JSON Interop & Prototype & FHIR (partial) & Not Specified & Patients, Doctors, Clinics & Dynamic \\
\hline
\cite{Pineda2022} & Two-layered (HLF + Hospital DB) & Permissioned-Consortium & Hyperledger Fabric & RAFT & Org-based AC (MSP + CA) & X.509, Org Policies & Prototype & National CA Policies & Colombian Regulation & Patients, Gov. Entities, Doctors & Dynamic \\
\hline
\cite{Stamatellis2020} & Two-layered (HLF + Encrypted DB) & Permissioned-Private & Hyperledger Fabric & RAFT & RBAC (MSP + CC) & Private Channels, CA/ACL & Prototype & Not Specified & Not Specified & Patients, Doctors & Static \\
\hline
\cite{Jakhar2024} & Two-layered (Ethereum + IPFS) & Permissionless-Public & Ethereum & PoW & RBAC (SC) & Encryption, Smart Contracts & Model & Not Specified & Not Specified & Patients, Doctors & Static \\
\hline
\cite{Boumezbeur2022} & Two-layered (Ethereum + Cloud) & Permissionless-Public & Ethereum & PoW & RBAC (SC) & AES + RSA, Auth Logs & Prototype & None & Not Specified & Patients, Doctors & Static \\
\hline
\cite{Rouhani2021} & Two-layered (Private BC + DB) & Permissioned-Private & Custom/Unspecified & Unspecified & ABAC (Policy Engine) & Policy Rules, Encrypted Log & Model & Not Specified & Not Specified & Patients, Doctors & Dynamic \\
\hline
\cite{Diaz2023} & Two-channel (HLF) & Permissioned-Consortium & Hyperledger Fabric & RAFT & RBAC (Private/Public Channels) & Channel Isolation, Audits & Prototype & Not Specified & Not Specified & Patients, Doctors, Insurers & Dynamic \\
\hline
\cite{Huang2022}& Two-layered (Ethereum + Hospital DB) & Permissionless-Public & Ethereum & PoW & RBAC (SC) & Encrypted DB, Audit Logs & Prototype & Not Specified & Not Specified & Patients, Doctors & Static \\
\hline
\cite{Pradhan2022} & Two-layered (HLF + Off-chain Server) & Permissioned-Private & Hyperledger Fabric & RAFT & RBAC (CC) & Private Channels, Smart Contract & Prototype & Not Specified & Not Specified & Patients, Clinics & Static \\
\hline
\cite{Mandarino2024} & Two-layered (Ethereum + IPFS) & Permissionless-Public & Ethereum & PoW & RBAC (SC) & SHA256, AES & PoC & None & Not Specified & Patients, Doctors & Static \\
\hline
\cite{Tahir2024b} & Two-layered (Ethereum + Cloud) & Permissionless-Public & Ethereum & PoW & RBAC (SC) & Encrypted Cloud, Auth Token & PoC & None & Not Specified & Patients, Doctors & Static \\
\hline
\end{tabular}
}

\vspace{0.2cm} \small
\noindent\textbf{HLF}: Hyperledger Fabric \quad
\textbf{BC}: Blockchain \quad
\textbf{CC}: Chaincode \quad
\textbf{SC}: Smartcontract \quad
\textbf{PbD}: Privacy by Design \quad
\textbf{ABE}: Attribute-Based Encryption 
\textbf{CA}: Certifcation Authority \quad

\textbf{MSP}: Membership Service Provider  \quad
\textbf{TEE}: Trusted Execution Environment \quad
\textbf{ZKP}: Zero-Knowledge Proof 
\textbf{VC}: Verifiable Credential \quad
\textbf{DID}: Decentralized Identifier
\end{sidewaystable*}

As for standard and regulatory compliance, only a handful of systems explicitly reference GDPR, HIPAA, or ISO standards. PbDinEHR~\cite{PbDinEHR} is an exception, offering structured mapping to ISO/IEC 27001 and 29100, while ACTION-EHR~\cite{ACTION-EHR} and ZeroTrustBlock~\cite{ZeroTrustBlock} embed consent and revocation capabilities aligned with GDPR. However, the overall lack of alignment between system features and concrete legal obligations suggests that privacy-by-design and regulation-aware engineering are still underdeveloped. Similarly, conformance with healthcare interoperability standards like HL7 FHIR is mentioned in only a few contributions, pointing to a need for greater integration of clinical exchange protocols to support EHR interoperability across platforms.

Finally, the comparative table illustrates that stakeholder modeling and consent mechanisms are largely simplified. Most systems involve only patients and healthcare providers, with limited roles for researchers, insurers, or public health authorities. Only a few contributions—such as those by Manoj et al.~\cite{Manoj2022} and Pineda Rincón~\cite{Pineda2022}—expand the ecosystem to include government regulators or verifiable research access. Consent mechanisms also vary: static models dominate, while dynamic and object-level consent—crucial for patient empowerment—are comparatively rare. This reflects a gap in stakeholder-inclusive, context-sensitive consent logic, which will be essential as EHR ecosystems scale.

These insights collectively point toward the need for more integrated, regulation-aware, and consent-centric designs in future BC-EHR systems, grounded in practical implementations that prioritize scalability, privacy, and institutional trust.

\section{Leveraging blockchain for secure healthcare data management}
\label{sec: 3}
Numerous works have explored the use of blockchain technology for healthcare data control, with a focus on the storage aspect. Due to the significant volume of health data, particularly high-resolution medical imaging, storage approaches have been categorized into two main axes: centralized and decentralized storage~\cite{pilkington2017can,wang2018secure,cao2019cloud,wang2019cloud,yang2018design}.

\subsection{Centralized storage}
\label{sec: 3.1}
Several studies have embraced a centralized data storage architecture using blockchain technology for managing access to and sharing health data. This approach offers simplicity in data management and exploitation. However, it also presents certain risks that can compromise the availability, integrity, and confidentiality of information. The centralization of data storage creates a potential vulnerability, where unauthorized access to the system can lead to unauthorized control over all the data. Moreover, this centralization introduces a "failure point" that can jeopardize system availability, whether due to physical disconnection or potential Denial of Service (DoS) attacks, resulting in the disruption of critical services.

While encryption is commonly used to protect health data within such a model, it does not eliminate the risk of data breaches. An illustrative example is the significant data leak in the American voting system, where a human error exposed voter data stored in a public cloud hosted by Amazon, as reported by CNN~\footnote{\url{https://money.cnn.com/2017/06/19/technology/voter-data-leaked-online-gop/index.html}}.

Another study by Dubovitskaya et al.~\cite{dubovitskaya2017secure}, in collaboration with Stony Brook University Hospital, proposed a cloud-based storage solution for health data related to cancer patients. They presented three scenarios: first aid, medical research, and a connected health system. This proposal involved a certification authority for generating asymmetric key pairs (public/private) for each user, which are used for encryption, decryption, and digital signatures. The patient's data stored in the cloud is encrypted using a symmetric key created by the patient. Data sharing occurs by exchanging the symmetric key, encrypted with the recipient's public key. In the event of a compromised symmetric key, the patient can generate a new one and re-encrypt their data in the cloud. However, this approach introduces several constraints that impact the system's optimal functioning, including:
\begin{itemize}
\item  The complexity of managing two pairs of asymmetric keys, one generated by a certification authority and the other native to the blockchain platform account.
\item The risk of compromising patient data due to their management of symmetric keys used for encrypting data in the cloud.
\item The lack of flexibility in the data-sharing process when multiple data entities are encrypted with the same symmetric key. Sharing one data entity without sharing others becomes impractical. However, using multiple keys exacerbates the challenge of key management.
\item The inefficiency of re-encrypting data whenever a patient suspects that their key has been compromised, which can slow down the system and introduce potential vulnerabilities.
\end{itemize}

These constraints highlight the need for alternative approaches to address the security and efficiency concerns associated with centralized data storage in blockchain-based healthcare systems.

\subsection{Decentralized storage}
\label{sec: 3.2}
Several recent studies have embraced the decentralized approach of using blockchain for managing health data stored in local databases. One notable project is MedRec, which builds upon the work of Azaria et al.~\cite{azaria2016medrec}. MedRec leverages blockchain technology to manage authorizations and access to medical data stored across network nodes. It is considered one of the pioneering functional prototypes in this domain. In MedRec, EHRs are stored in network nodes, while smart contracts deployed on the Ethereum blockchain facilitate active interaction between them. The platform utilizes a public blockchain and employs a Proof of Work (PoW)\footnote{Proof of Work: A consensus mechanism where the miners involved in the validation mechanism will have to perform a mathematical calculation that is difficult to find but easy to prove} validation mechanism, involving transaction costs and mining processes for account management.

While MedRec is currently operational in Boston (USA), we have identified certain architectural weaknesses that could hinder its seamless operation, including:
\begin{itemize}
\item  The requirement for patient validation before adding a record to the blockchain. This approach poses a disadvantage as patients may not actively validate such requests, potentially causing delays in adding records. To address this, health actors should be able to create and add records for a patient at any time without patient validation, while access to those records by other actors should still require patient authorization.
\item  The assumption that patient workstations must have a database, which may not always be feasible, as patients may use small devices that cannot support such applications.
\end{itemize}

Addressing these weaknesses will be crucial to ensuring the effective implementation and utilization of decentralized storage solutions in blockchain-based healthcare systems.

\section{A secure and distributed architecture for EHR sharing and access control}
\label{sec: 4}

\begin{figure}[h]
\centering
\includegraphics[width=0.45\textwidth]{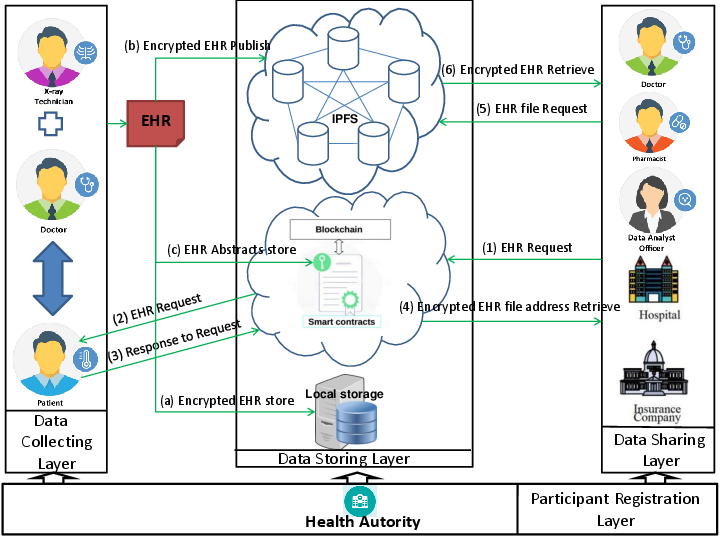}
\caption{Overall view of the proposed architecture}
\label{fig:1}       
\end{figure}

To meet the increasing need for sharing EHRs between health providers, especially during a pandemic such as COVID-19, this work proposes a reliable, secure, and extensible solution. Our proposal is a reusable architecture (Figure~\ref{fig:1}) in the form of a set of protocols that offer services to different health providers wishing to share EHRs securely. This architecture is based on a private blockchain to manage the sharing and access control of EHRs while keeping them stored in a secure and decentralized manner using IPFS.  

The operation of the private blockchain in our proposed architecture involves the establishment of a consortium comprising relevant healthcare stakeholders, such as healthcare providers, institutions, and health authorities. The selection of consortium members is based on a collaborative approach, taking into account the specific requirements of the healthcare ecosystem and the need for a diverse yet trusted network. This approach ensures that the private blockchain is operated and governed by entities with a vested interest in the secure and interoperable management of EHRs. The merits of our consortium-based private blockchain approach over designs involving trusted third parties or public key infrastructures lie in the decentralized and transparent nature of the blockchain. By distributing control among consortium members and leveraging smart contracts, we enhance security, reduce the risk of a single point of failure, and empower patients with direct control over their health information. 
 
\subsection{Overview of the proposed architecture}
\label{sec: 4.1}
The proposed architecture (Figure~\ref{fig:1}) is based on smart contracts deployed within a private blockchain and IPFS. Four main actors are involved in this architecture, each with specific roles and credentials, namely the patient, the health actor, the health provider, and the Health Authority, by creating and exchanging Electronic Health Records.
\begin{itemize}
    \item \textbf{Patient:} He is the person who requests the services offered by the health provider. He is the main actor on the platform.
    \item \textbf{Health Actor:} He is the representative of health service providers; he interacts with the patient to give him the necessary services. He can be a doctor, nurse, health insurance professional, laboratory technician, etc. The health actor must work in an establishment recognized by the Health Authority.
    \item \textbf{Health Provider:}   It is the institution where the health actor exercises his profession; it can be a hospital, a medical office, a laboratory, or a medical insurance office. Each health provider should have a data storage database. Otherwise, another health center responsible for storing EHRs should be specified.
    \item \textbf{Health Authority:} A shared health system does not accept anonymity under any circumstances. That's why we suggest creating a trusted institution capable of recognizing the authenticity of the participants in the system. All participants will have to submit their requests to join the system, and it is up to this authority to decide the fate of these requests. This authority is also responsible for deploying smart contracts on the blockchain.
\end{itemize}

Actors interact with each other through one of the four layers of the architecture, namely 1) Participant enrollment layer responsible for user registration, 2) Data collection layer responsible for EHR creation, 3)  Data storage layer responsible for local and distributed EHR storage, and 4)  Data sharing layer responsible for EHRs access. 

Regarding the service rendered by a health actor, a digital document is produced for the benefit of the patient, such as a prescription or medical imaging. It is commonly called an EHR, which may be seen as the currency to be exchanged. To allow a patient's EHR secure sharing between multiple health providers and practitioners, five (05) protocols are defined in our architecture. These protocols ensure almost all functionalities expected in a sharing EHR system. 
\begin{enumerate}
    \item \textbf {Registration Protocol (EHR-RP)} which ensures identification and registration of participants, 
    \item \textbf {EHR Publication Authorization Protocol (EHR-PUB-AUTH)} which controls access to health data, 
    \item \textbf {EHR Publishing Protocol (EHR-PUB)} which provides health data to the system, 
    \item \textbf {EHR Integrity Verification Protocol (EHR-IV)}, which is responsible for protecting the EHR from any integrity compromise, 
    \item \textbf {EHR Accessing Protocol (EHR-ACCES)} which ensures secure data sharing.
\end{enumerate}

We should notice that communications between healthcare actors aim to exchange EHR and can be achieved on blockchain or off-blockchain, depending on the sensitivity of the exchanged data. 
\begin{itemize}
    \item \textbf{On-blockchain communications:} They are communications that pass through the blockchain. They are controlled by smart contracts.
    \item \textbf{Off-blockchain communications:} They are communications carried out directly between the actors of the system, without involving the blockchain. They are controlled by the actors using already obtained data through On-blockchain communications.
\end{itemize}

\begin{figure}[h]
\centering
\includegraphics[width=0.45\textwidth]{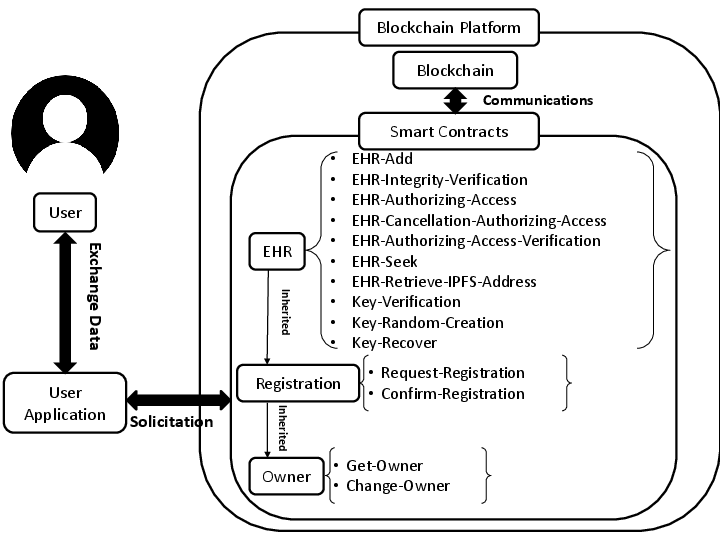}
\caption{Smart contracts of our architecture}
\label{fig:2}       
\end{figure}

\subsection{Smart Contracts}
\label{sec: 4.2}
In the proposed architecture, three (03) smart contracts are deployed, namely "Owner", "Registration", and "EHR", which are responsible for managing On-blockchain communications. As shown in Figure \ref{fig:2}, users communicate with their interface application using the system protocols. A user in the system can invoke the functions defined in the three smart contracts for which they are authorized. In the following, we give a summary of the main functions of the three (03) smart contracts.


\subsubsection*{A. The smart contract "Owner"}
\label{sec: 4.3}
This contract allows the identification, recovery, or change of the owner of all the smart contracts. By default, the owner is the user who deployed it to the blockchain for the first time. In our case, the owner is the Health Authority. The two main functions of this smart contract are defined as follows:
\begin{enumerate}
\item \textbf{Get-Owner:}
This function has the task of saving and returning the blockchain account that owns the smart contract.
\item \textbf{Change-Owner:}
The purpose of this function is to change the owner of the smart contract. It is mainly useful in the case of changing the owner account of smart contracts without stopping their operation. Notice that this function can be invoked only by the Health Authority.
\end{enumerate}

\subsubsection*{B. The smart contract "Registration"}
\label{sec: 4.4}
The "Registration" smart contract inherits from the smart contract "Owner" all its variables and functions. It is responsible for the registration phase of the overall system process. It contains two (02) main functions, namely:
\begin{enumerate}
\item \textbf{Request-Registration:}
The "Request-Registration" function is responsible for saving the accounts wishing to join the platform, pending the decision of the Health Authority concerning their validation.
\item \textbf{Confirm-Registration:}
This function is used exclusively by the Health Authority (the owner of the smart contract) to validate, or not, a given request.
\end{enumerate}

\subsubsection*{C. The smart contract "EHR"}
\label{sec: 4.5}
It inherits from the "Registration" smart contract and, by transitivity, from the "Owner" smart contract. It is responsible for all processing, handling, and verification operations carried out on the EHRs within the platform. It has eight (08) main functions, namely:

\begin{enumerate}
\item \textbf{EHR-Add:}
This function ensures the addition of EHRs to the blockchain and the execution of all the necessary checks for the accomplishment of this operation.

\item \textbf{EHR-Integrity-Verification:}
This function is responsible for verifying the integrity of an EHR entered as an input against its copy recorded in the blockchain.

\item \textbf{EHR-Authorizing-Access:}
This function allows the creation of an authorization to consult a given EHR for the benefit of a given actor.

\item \textbf{EHR-Cancellation-Authorizing-Access:}
This function allows the cancellation of an existing authorization in the smart contract.

\item \textbf{EHR-Authorizing-Access-Verification:}
This function checks the existence of an authorization stored in the system with an identifier entered as input.

\item \textbf{EHR-Seek:} This function allows searching for a specific EHR.

\item \textbf{EHR-Retrieve-IPFS-address:} This function returns the IPFS address of a given EHR.

\item \textbf{Key-Verification:}
This function is responsible for verifying the authenticity of a key entered as input compared to that stored in the blockchain with the same identifier.

\item \textbf{Key-Random-Creation:}
The mission of this function is to create a random key using several random system parameters, such as:
\begin{itemize}
    \item System time;
    \item Identifier of the active Hyperledger account;
    \item Variable incremented each time the function is used;
\end{itemize}

\item \textbf{Key-Recover:}
This function allows for the retrieval of an authorization key stored in the blockchain.
\end{enumerate}

\section{Main protocols of the proposed architecture}
\label{5}
In the following, we give an in-depth description of the main procedures and protocols of the architecture to ensure a secure and distributed architecture for EHR sharing and access control.

\subsection{Participants registration}
\label{sec: 5.1}
The health authority must manage the identities of all actors separately in order to avoid possible collisions that may occur when adding new entities to the system. 

The registration of the participants is done through the Registration Protocol\textbf{ (EHR-RP)}, using the smart contract "Registration", for the validation and registration of the new system participants. 

Here, we assume that every actor in the system (Patients, Doctors, Operators, etc.) will have an account on the blockchain because of successful registration. A private blockchain has to be deployed over the main health providers and maintained by the Health Authority. Then, every entity can have a dedicated user application preloaded with a blockchain credential. This can be a web or mobile application for citizens and an integrated application within the health provider information system for practitioners.


\begin{figure}[h]
\centering
\includegraphics[width=0.45\textwidth]{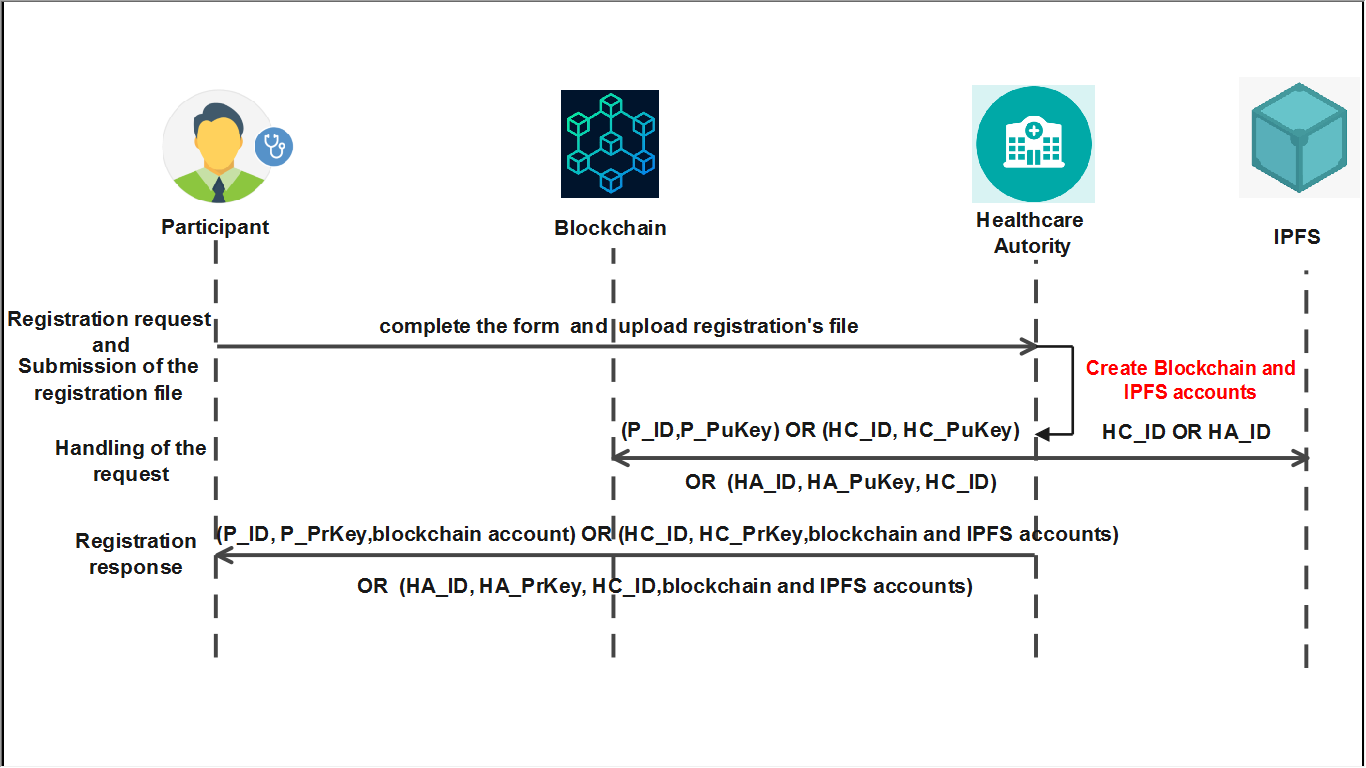}
\caption{Registration protocol (only for health centers and health actors, accounts for patients are automatically created by the Health Authority)} 
\label{RP}
\end{figure}

\subsubsection*{Registration Protocol (EHR-RP)} 
\label{sec: 5.2}
Access to the platform is guaranteed for everyone, but participants first have to prove their identity and their role to the health authority. To implement this principle, the \textbf{EHR-RP} (Figure \ref{RP}) consists of two main interfaces. The first one is used by everyone, in non-blockchain mode, to submit their registration requests by filling out forms containing their identity and professional information, as well as information related to the role they are applying for within the platform. This request is accompanied by the transfer of the application file (birth certificate, identity or professional card, degree, administrative status, etc.). 

The second interface, which ensures the registration and management of the participants in the blockchain, is used exclusively by the health authority, the only entity entitled to accept or refuse the registration request, after verification of the file submitted by the applicant. Passing through the RP protocol is mandatory to allow interaction with the system for all stakeholders.

\subsection{EHR creation and publishing}
\label{sec: 5.3}
The patient obtains an EHR after interaction with a healthcare provider, and a copy of this document is immediately stored locally via the health center's information system. In order to preserve the confidentiality of the patient's medical data, it is recommended to encrypt the document with a secret key generated by the local health center's system before it is stored in the dedicated databases. Only authorized users have access to the stored information. To provide access to the patient's EHR to other stakeholders, we propose that the document should be published in the IPFS network. The IPFS network offers the possibility to store large patient EHRs. This publication requires prior authorization from the patient. 

\subsubsection*{A. EHR publishing authorization}
\label{sec: 5.4}
Once an EHR is created as a result of patient interaction with a health actor, the latter asks the patient to validate the EHR publication authorization through the \textbf{EHR-PUB-AUTH} protocol. In this protocol, the health actor invokes the \textbf{EHR-Authorizing-Access} function of the "EHR" smart contract, and the patient validates the publication authorization. This will generate and store the publication authorization in the blockchain. The health actor receives a notification of the publishing authorization issued by the patient, which must include the following identifiers.
\begin{itemize}
\item $AU-ID$: Authorization identifier
\item $HA-ID$: Health Actor identifier 
\item $PT-ID$: Patient identifier
\item $EHR-ID$: EHR identifier
\item $HC-ID$: Health provider identifier
\end{itemize}


\begin{figure}[h]
\centering
\includegraphics[width=0.45\textwidth]{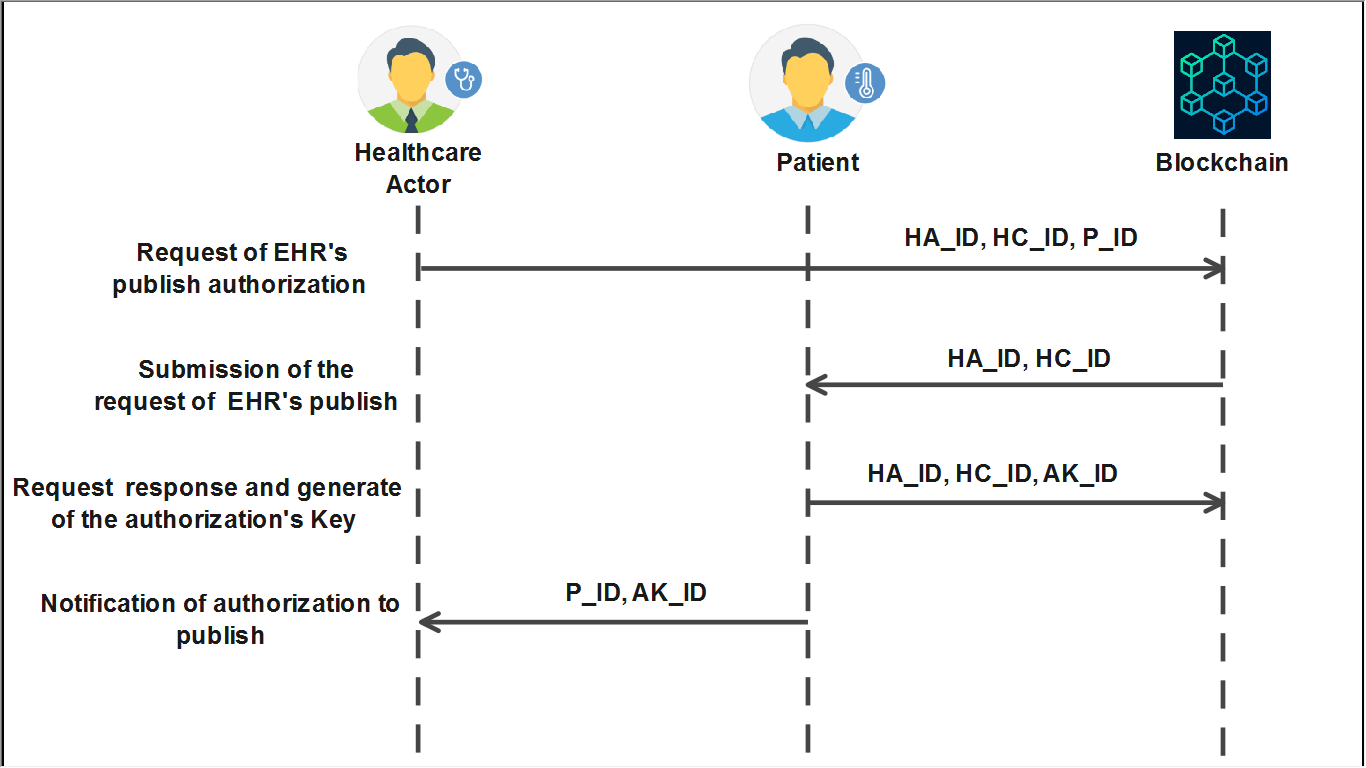}
\caption{EHR publishing authorization protocol} 
\label{EHR-AP}
\end{figure}

\begin{figure}[h]
\centering
\includegraphics[width=0.45\textwidth]{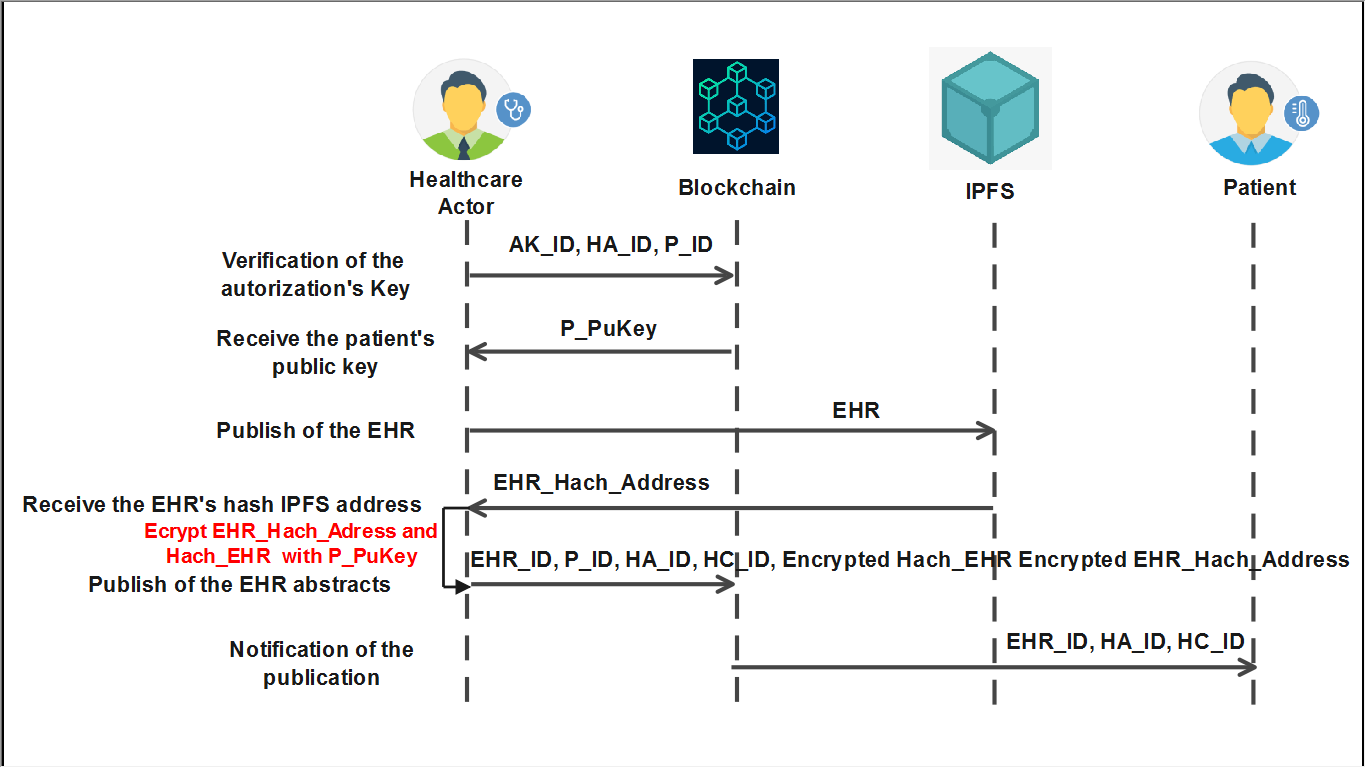}
\caption{EHR publishing protocol} 
\label{EHR-publishing}
\end{figure}

The main steps for executing the \textbf{EHR-PUB-AUTH} protocol are shown in Figure~\ref{EHR-AP}. The EHR publication operation, and its summary, will be triggered only after the health actor receives the publication authorization, generated for the benefit of the patient concerned. Indeed, the EHR will be published in the IPFS network, and its summary will be stored in the blockchain. This operation is ensured by the EHR-PUB protocol (Figure~\ref{EHR-publishing}), using the dedicated functions of the "EHR" smart contract, as it will be explained in the next subsection.

\subsubsection*{B. EHR publishing}
\label{sec: 5.5}
Once the health actor has received the authorization to publish the EHR, and before proceeding to its publication, the \textbf{EHR-Authorizing-Access-Verification} function of the "EHR" smart contract will be executed. After confirming the authenticity of the authorization and receiving the patient's public key, the EHR file will be encrypted with a randomly generated secret key using the AES-256 symmetric encryption technique. This file will be sent to the IPFS network for publication, and in return for this step, the health actor will receive the hash address of the encrypted EHR. The returned address and secret key will be encrypted by the patient's public key.

By invoking the "\textbf{EHR-Add}" function of the EHR smart contract, the following EHR-related data will be stored in the blockchain, namely: 
\begin{itemize}
    \item $EHR-ID$: The EHR identifier in the health center.
    \item $PT-ID$: The identifier of the patient concerned by the EHR.
    \item $HA-ID$: The identifier of the health actor producing the EHR.
    \item $HC-ID$: The identifier of the health center where the EHR was produced.
    \item $EHR-TP$: The type of EHR (a prescription, medical imaging, or other).
    \item $KWD$: The list of keywords for indexing the EHR established by the health actor.
    \item $EHR-HS$: The EHR hash encrypted with the patient's public key.
    \item $EHR-IPFS$: The hash address of the EHR encrypted by the patient's public key.
    \item $SK$: The secret key is encrypted by the patient's public key.
\end{itemize}
    

At the end of this operation, the patient will be notified about the publication of his or her EHR, and the reference of the EHR will be listed in the EHR publication list in his or her user account.

\subsection{EHR sharing}
\label{sec: 5.6}
This layer ensures the secure sharing of EHRs by enabling a health actor to perform the following steps. First, the identification of the EHR to be consulted, notably through a keyword search in the EHR summaries stored in the blockchain, using the "\textbf{EHR-Seek}" function of the EHR smart contract. Then, the health actor has to send a request for authorization to consult the EHR addressed to the patient concerned via the \textbf{EHR-ACCES} protocol. If access is granted, the EHR's IPFS address will be retrieved via the EHR Address Recovery function of the EHR smart contract. Finally, the health actor can access the EHR directly from the IPFS, after decrypting the EHR hash address and querying the IPFS to download the requested EHR.  Optionally, a verification of the integrity of the EHR via the \textbf{EHR-IV} protocol can be done.

\subsubsection*{A. EHR access request}
\label{sec: 5.7}
\begin{figure}[h]
\centering
\includegraphics[width=0.45\textwidth]{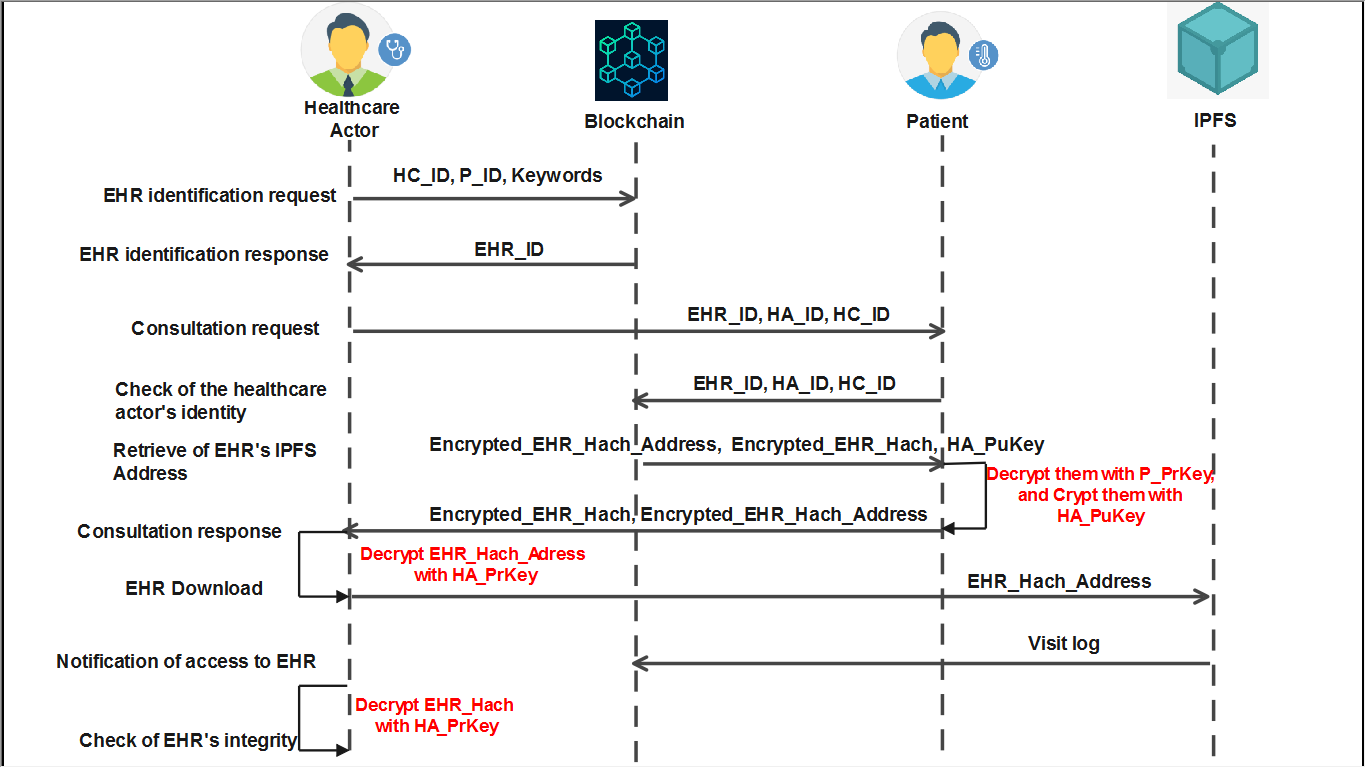}
\caption{EHR accessing protocol} 
\label{EHR-ACP}
\end{figure}

Access to EHRs is prohibited by default. The \textbf{EHR-ACCES} protocol gives exclusivity to the patient who owns the EHR to authorize a health actor requesting access (Figure \ref{EHR-ACP}). Once a health actor has retrieved the EHR identifier, he or she will request the owner (patient) to grant access to his or her EHR. The patient will receive the request containing the health actor identifier and its public key, and the requested EHR identifier.
Through the function "\textbf{EHR-Retrieve-IPFS-address}" of the "EHR" smart contract, the patient will retrieve the EHR's hash address and its Hach, previously encrypted by his public key. He will decrypt them with his private key and encrypt them with the public key of the health actor. At the end of this operation, the requester will receive the notification of the authorization, containing the IPFS address, the secret key, and the hash of the EHR encrypted by his public key.


\subsubsection*{B. EHR Integrity Verification}
\label{sec: 5.8}
Every user of the system can benefit from the services of this protocol to check the EHR integrity before being used. Thus, the system calculates the hash of the introduced EHR and compares it to the one stored in the blockchain, and then responds to whether the EHR is intact or not.

\begin{enumerate}
\item An actor introduces an EHR for integrity verification by the system.
\item The system uploads the EHR reference and the hash. After verification, the system responds either with "OK" if the EHR is authentic or with "NO" if it is corrupted.
\end{enumerate}

\subsubsection*{C. EHR accessing}
\label{sec: 5.9}

In this step, once the requester has retrieved the IPFS address of the EHR, it will decrypt it by its private key, and the requester queries the IPFS to retrieve the patient's EHR. In this phase, no verification is required for the completion of the EHR-ACP protocol.

It should be noted that not all these operations are visible to the clinician working with its health provider's information system. As we will see in the next section, all operations involving the blockchain are implemented in the blockchain ecosystem and are just remotely invoked by end-user applications.

Table~\ref{tab1} explains the smart contract functions used by the different platform protocols to perform their tasks involving On-blockchain communications.

\begin{table}[t]
\centering
\footnotesize
\caption{Smart Contracts Functions per Protocol}
\label{tab1}
\renewcommand{\arraystretch}{0} 
\begin{tabular}{|>{\centering\arraybackslash}m{0.03\linewidth}|>{\centering\arraybackslash}m{0.25\linewidth}|>{\centering\arraybackslash}m{0.55\linewidth}|}
\toprule
\textbf{} & \textbf{Protocols} & \textbf{Triggered Functions / Smart Contracts} \\
\midrule
01 & EHR-RP  & 
\begin{itemize}[topsep=0pt, partopsep=0pt, itemsep=0pt, parsep=0pt, leftmargin=*]
  \item Request-Registration / Registration
  \item Confirm-Registration / Registration
  \item Get-Owner / Owner
\end{itemize} \\
\midrule
02 &  EHR-PUB-AUTH &
\begin{itemize}[topsep=0pt, partopsep=0pt, itemsep=0pt, parsep=0pt, leftmargin=*]
  \item EHR-Authorizing-Access / EHR
  \item EHR-Cancellation-Authorizing-Access / EHR
\end{itemize} \\
\midrule
03 & EHR-PUB  &
\begin{itemize}[topsep=0pt, partopsep=0pt, itemsep=0pt, parsep=0pt, leftmargin=*]
  \item EHR-Add / EHR
\end{itemize} \\
\midrule
04 & EHR-IV  &
\begin{itemize}[topsep=0pt, partopsep=0pt, itemsep=0pt, parsep=0pt, leftmargin=*]
  \item EHR-Integrity-Verification / EHR
\end{itemize} \\
\midrule
05 &  EHR-ACCESS &
\begin{itemize}[topsep=0pt, partopsep=0pt, itemsep=0pt, parsep=0pt, leftmargin=*]
  \item EHR-Authorizing-Access-Verification / EHR
  \item EHR-Integrity-Verification / EHR
  \item Key-Verification / EHR
  \item Key-Random-Creation / EHR
  \item Key-Recover / EHR
\end{itemize} \\
\hline
\end{tabular}
\end{table}

\section{Experimental evaluation} 
\label{sec: 6}
A prototype of our architecture has been implemented to evaluate the proposed protocols. Regarding the blockchain, the Hyper-Ledger platform was chosen to implement our proposal, given its suitability for private deployment.

\subsection{Illustrative use cases}
Several use cases have been tested on the developed prototype by simulating normal and malicious scenarios. To understand the functionalities offered by our solution, we will focus on an illustrative scenario of legitimate and correct use that involves all the protocols of the system. The actors participating in this scenario are :
\begin{itemize}
    \item Health Authority (HA);
    \item 01 Patient (PT);
    \item 02 Health Actors: Doctor 1 (DOC1), Doctor 2 (DOC2);
    \item 02 Health Centers: Hospital 1 (HP1), Hospital 2 (HP2).
\end{itemize}

Here, we used JavaScript code to simulate all actors in this scenario. The steps for setting up this scenario are as follows:

\subsubsection*{Step 1:  Initialization}
\label{sec: 6.1}
\begin{itemize}
    \item The deployment, by the HA, of smart contracts in the blockchain and recovery of its API;
    \item Integration, by PT, DOC1, DOC2, HP1, and HP2, of the API of the deployed smart contracts into their local applications. 
\end{itemize}

\subsubsection*{Step 2:  Registration}
\label{sec: 6.2}
\begin{itemize}
    \item Registration request from PT, DOC1, DOC2, HP1, and HP2 (using EHR-RP);
    \item After checking their files, HA validates the registration requests of PT, DOC1, DOC2, HP1, and HP2 and sends acceptance notifications.
\end{itemize}

\subsubsection*{Step 3:  First medical consultation}
\label{sec: 6.3}
\begin{itemize}
    \item PT consults his attending physician, DOC1, at HP1;
    \item DOC1 administers two medical prescriptions, EHR1 and EHR2, which will be stored in the HP1 databases (using EHR-AUTH and EHR-PUB).
\end{itemize}

\subsubsection*{Step 4: Second medical consultation - integrity check}
\label{sec: 6.4}
\begin{itemize}
    \item PT, far from home, consults DOC2 at HP2. DOC2 does not know the patient and asks him for his medical records, but PT has only EHR1.
    \item DOC2 recovers EHR1 and ensures its integrity (via EHR-IV).
\end{itemize}

\subsubsection*{Step 5: Second medical consultation - EHR2 accessing authorization}
\label{sec: 6.5}
\begin{itemize}
    \item Due to the absence of EHR2 in the medical files presented by PT, DOC2 asks him for the authorization to consult his EHR2 via the Platform.
    \item PT, via its interface (mobile App), authorizes DOC2 to consult EHR2.
    \item DOC2 receives a notification providing him with the authorization to consult EHR2 domiciled in HP1. Then DOC2 consults EHR2 by performing the rest of the actions included in the EHR-ACCESS.
\end{itemize}

\subsubsection*{Step 6: Second medical consultation - EHR3 adding}
\label{sec: 6.6}
\begin{itemize}
    \item After consulting his medical file, DOC2 administers the EHR3 medical prescription to PT using the EHR-PUB protocol.
\end{itemize}

\subsubsection*{Step 7: Cancellation of authorization to consult EHR2}
\label{sec: 6.7}
\begin{itemize}
    \item Cancellation, by PT, of the authorization to consult the EHR2 granted to DOC2 since the latter is not his attending physician.
\end{itemize}

In addition to the described scenario, several others were tested, including those that show improper use of the system. Table~\ref{tab2} shows the interaction of the system with the incorrect use of the platform.

\begin{table}[h!]
\centering
\footnotesize
\caption{System Interaction Against Improper Use of the Platform}
\label{tab2}
\renewcommand{\arraystretch}{0}
\begin{tabular}{|>{\centering\arraybackslash}m{0.03\linewidth}|>{\centering\arraybackslash}m{0.45\linewidth}|>{\centering\arraybackslash}m{0.35\linewidth}|}
\toprule
\textbf{} & \textbf{Cases} & \textbf{System Reactions} \\
\midrule
01 & Usurpation of the function of the Health Authority to validate registration requests of new users or to cancel the validation of user accounts. & 
\begin{itemize}
\item Exit;
\item Cancellation of account validation;
\item Health Authority notification.
\end{itemize} \\
\midrule
02 & Attempt to consult an EHR via a canceled authorization. &
\begin{itemize}
\item Exit;
\item User notification.
\end{itemize} \\
\midrule
03 & No prior registration request using the presented account. & 
\begin{itemize}
\item Exit;
\item User notification.
\end{itemize} \\
\midrule
04 & The presented account is not validated by the Health Authority & 
\begin{itemize}
\item Exit;
\item User notification.
\end{itemize} \\
\midrule
05 & Use of an account canceled by the Health Authority & 
\begin{itemize}
\item Exit;
\item User notification.
\end{itemize} \\
\midrule
06 & Using EHR-PUB protocol for adding EHR without having the credentials (a nurse who adds a medical prescription) & 
\begin{itemize}
\item Exit;
\item Health Authority notification.
\end{itemize} \\
\midrule
07 & Malicious attempt to authorize consultation of another patient's EHR & 
\begin{itemize}
\item Exit;
\item Health Authority notification.
\end{itemize} \\
\midrule
08 & Attempt to consult an EHR without authorization & 
\begin{itemize}
\item Exit;
\item Health Authority notification.
\end{itemize} \\
\midrule
09 & Usurpation of user identity by presenting with an account different from that declared during the registration phase & 
\begin{itemize}
\item Exit;
\item Health Authority notification.
\end{itemize} \\
\bottomrule
\end{tabular}
\end{table}

\subsection{Hyperledger Network Performance Tests}

\begin{figure}[h]
\centering
\includegraphics[width=0.5\textwidth]{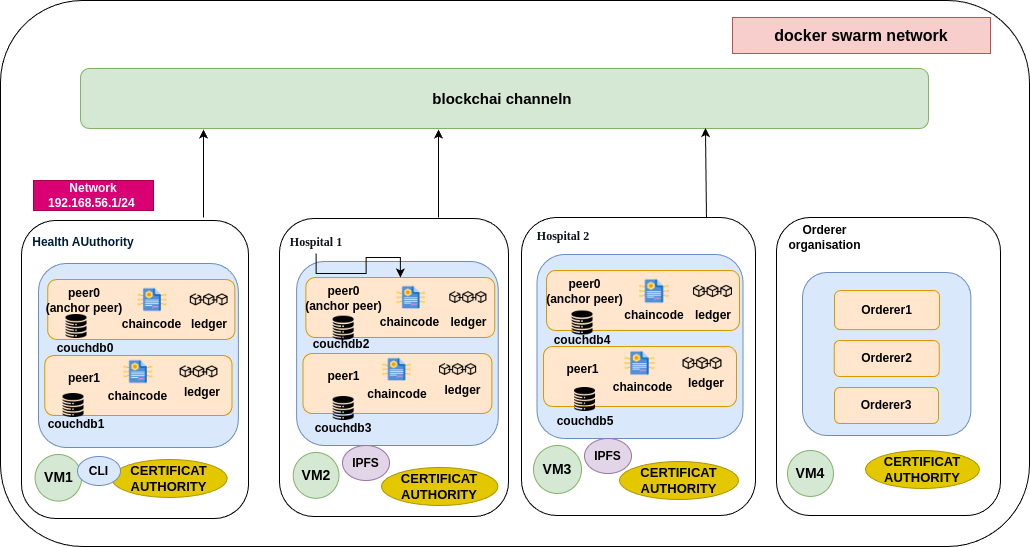}
\caption{Hyperledger Network deployed.}
\label{fig:Hyperledger}
\end{figure}

This section presents the results of performance tests conducted on a Hyperledger Fabric network deployed on four Ubuntu 22.04 virtual machines as depicted in Figure \ref{fig:Hyperledger}. The tests were performed using the Hyperledger Caliper tool, implemented in JavaScript, on a host machine equipped with an AMD Ryzen 7 5800U processor and 16 GB of RAM. These tests aim to evaluate different aspects of network performance, including the impact of timeout, transaction latency, and network traffic, as well as  CPU and RAM consumption, depending on parameters such as the number of transactions and ledger size. 

Notice that we distinguish between two types of requests: read and write. \textit{Read} requests need only access to the ledger; however, \textit{Write} requests need to modify the distributed ledger and will involve the execution of a transaction. 

\subsubsection{Impact of TPS and Timeout  on the Number of Successful Transactions}

This first test examines the effect of timeout (i.e., the maximum delay allowed to execute a transaction; otherwise, it is canceled) and the number of Transactions Per Second (TPS) on the successful transaction rate, a critical parameter for ensuring the responsiveness of our platform. We set the total number of transactions to 2000 for each test. 

For write requests, the number of transactions per second was varied between 10 and 50, with a timeout between 1 and 60 seconds. The results indicate that a short timeout combined with a high transaction rate reduces the number of successful transactions. 

For read requests, we varied the number of transactions per second between 100 and 600, with a timeout between 1 and 15 seconds. A similar trend was observed: a low timeout with a high rate decreases the number of successful transactions.

\begin{figure}[h!]
\centering
\begin{subfigure}{0.5\textwidth}
\centering
\includegraphics[width=1\textwidth]{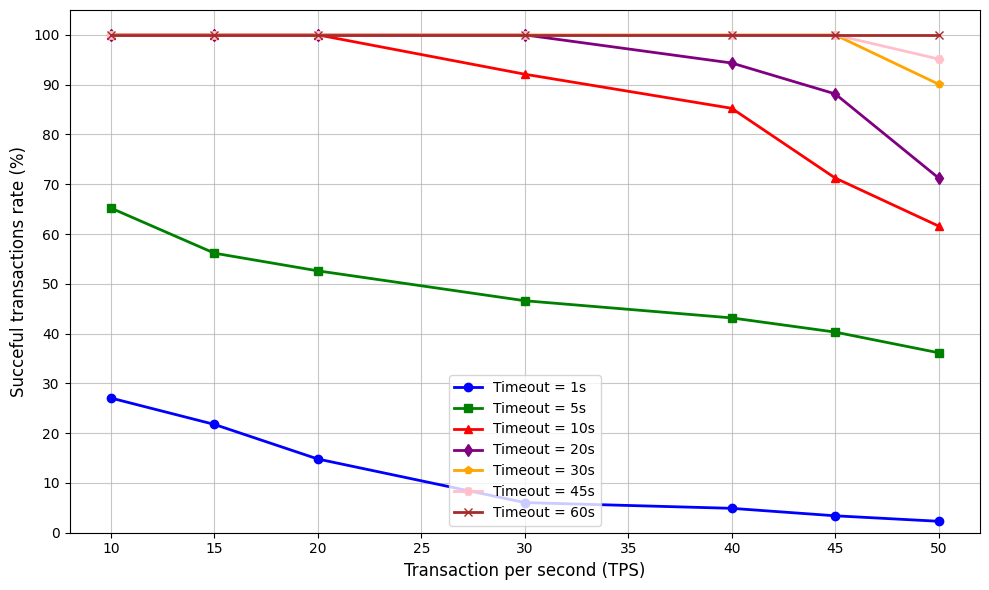}
\caption{Write requests}
\label{fig:test1_read}
\end{subfigure}
\par\bigskip 
\begin{subfigure}{0.5\textwidth}
\centering
\includegraphics[width=0.97\textwidth]{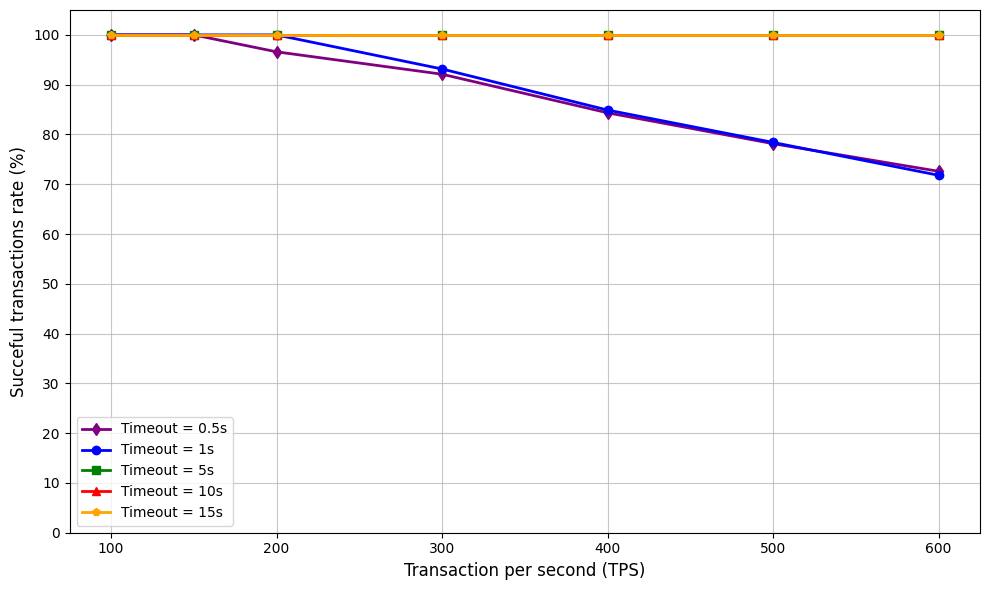}
\caption{Read requests}
\label{fig:test1_write}
\end{subfigure}
\caption{Number of successful transactions based on timeout and transactions per second.}
\label{fig:test1}
\end{figure}

As shown in Figure~\ref{fig:test1}, our network can handle up to 600 read transactions per second with a 5-second timeout and up to 45 write transactions per second with a 30-second timeout. This difference is explained by the fact that writes require ledger modification and ordering by the \emph{orderers}, unlike reads, which directly access a peer's ledger.

Based on this, we adjusted the parameters for subsequent tests, adopting a timeout of 30 seconds and a rate of 40 transactions per second for write requests. Also, we chose a timeout of 5 seconds and a 200 transactions per second rate for reads.


\subsubsection{Transaction Latency Based on Ledger Size}

This test evaluates the impact of ledger size on transaction latency, an important aspect for our platform, where the ledger can grow with the addition of EHRs. Two functions are tested: \textbf{EHR-Add} (write, 40 transactions per second, 1000 transactions per test) and \textbf{EHR-Seek} (read, 200 transactions per second, 1000 transactions per test).

\begin{figure}[h]
\centering
\includegraphics[width=0.5\textwidth]{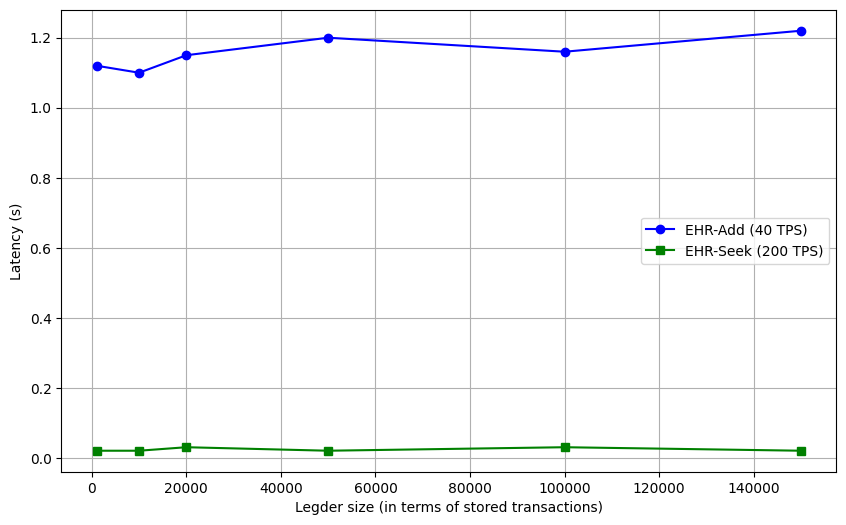}
\caption{Transaction latency based on ledger size.}
\label{fig:test2}
\end{figure}

The ledger size is progressively increased up to 150,000 records. The results (Figure~\ref{fig:test2}) show stable latency: for writes, it oscillates between 1.1 and 1.2 seconds, and for reads, it varies between 0.01 and 0.03 seconds, regardless of ledger size. This stability suggests that Hyperledger Fabric effectively manages the increasing ledger size in our solution.


\subsubsection{Transaction Latency Based on the Number of Transactions}

This test analyzes the effect of the number of transactions on the latency by varying the number of transactions for the optimal TPS. Notice that we empty the ledger before each iteration to start with a blank blockchain. We used the functions \textbf{EHR-Add} (write, 40 transactions per second) and \textbf{EHR-Seek} (read, 200 transactions per second), varying the total number of transactions from 1000 to 8000.

\begin{figure}[h]
\centering
\includegraphics[width=0.5\textwidth]{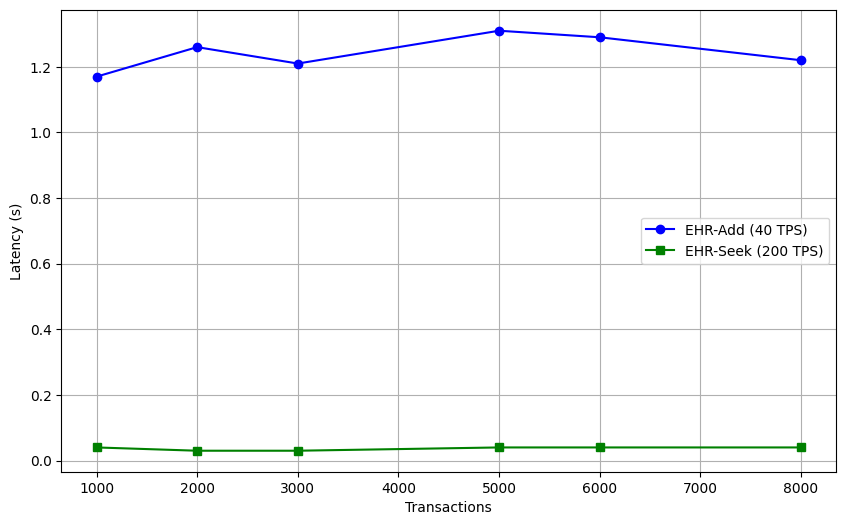}
\caption{Transaction latency based on the number of transactions.}
\label{fig:test3}
\end{figure}

The results (Figure~\ref{fig:test3}) indicate stable latency: for writes, it fluctuates between 1.19 and 1.35 seconds, and for reads, between 0.01 and 0.03 seconds, regardless of the number of transactions. This consistency shows the network's ability to maintain its performance in the face of increasing load for the optimal TPSs.


\subsubsection{Network Traffic Analysis}

This test measures the incoming (\textit{in}) and outgoing (\textit{out}) network traffic generated by the execution of 2000 transactions collected from network nodes. We examined both types of requests: write (\textit{EHR-Add}, data addition) and read (\textit{EHR-Seek}, data retrieval).

\begin{figure}[h]
\centering
\includegraphics[width=0.5\textwidth]{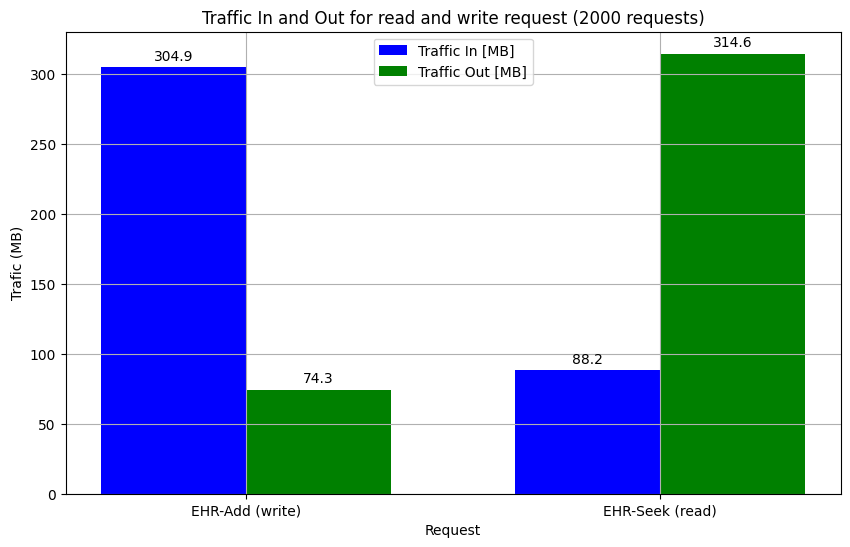}
\caption{Incoming and outgoing network traffic.}
\label{fig:test4}
\end{figure}

The results (Figure~\ref{fig:test4}) reveal notable differences: \textit{EHR-Add} requests generate higher incoming traffic than outgoing traffic due to data sent to modify the ledger, while \textit{EHR-Seek} requests produce more outgoing traffic, reflecting the responses returned to the requests. These observations are crucial for optimizing network management in our solution.


\subsubsection{CPU Consumption Based on TPS}

This test evaluates CPU consumption based on transactions per second (TPS), with 2000 transactions and a 30-second timeout. For writes (\textit{EHR-Add}), the TPS  varied from 5 to 50, and for reads (\textit{EHR-Seek}), from 50 to 500.

\begin{figure}[h]
\centering
\includegraphics[width=0.45\textwidth]{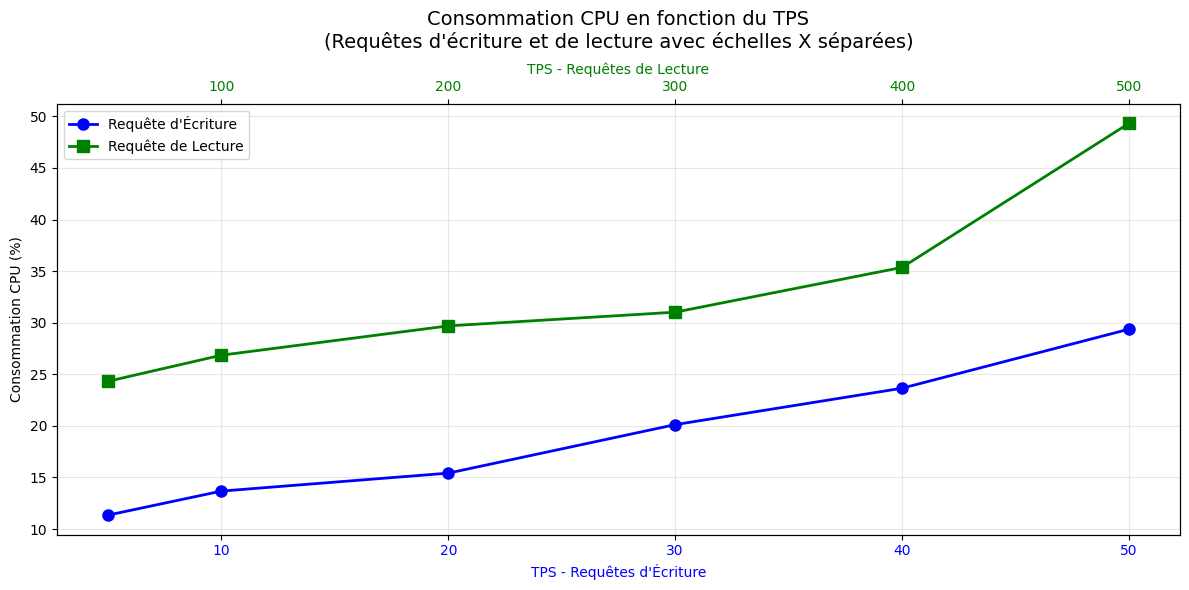}
\caption{CPU consumption based on TPS.}
\label{fig:test5}
\end{figure}

The results (Figure~\ref{fig:test5}) show a proportional increase in {CPU} consumption with {TPS}: for read requests, it increases from 10.3\% to 29.3\%, and for write requests, from 24.3\% to 49.4\%. This trend reflects the increasing allocation of CPU resources to process a higher volume of transactions, a factor to consider for our platform during peak usage. These observations allow us to anticipate hardware resource needs and optimize node configuration to maintain stable performance under heavy load. Measurements were performed on peer and orderer containers, with real-time monitoring of CPU metrics.


\subsubsection{RAM Consumption Based on TPS}

\begin{figure}[h]
\centering
\includegraphics[width=0.45\textwidth]{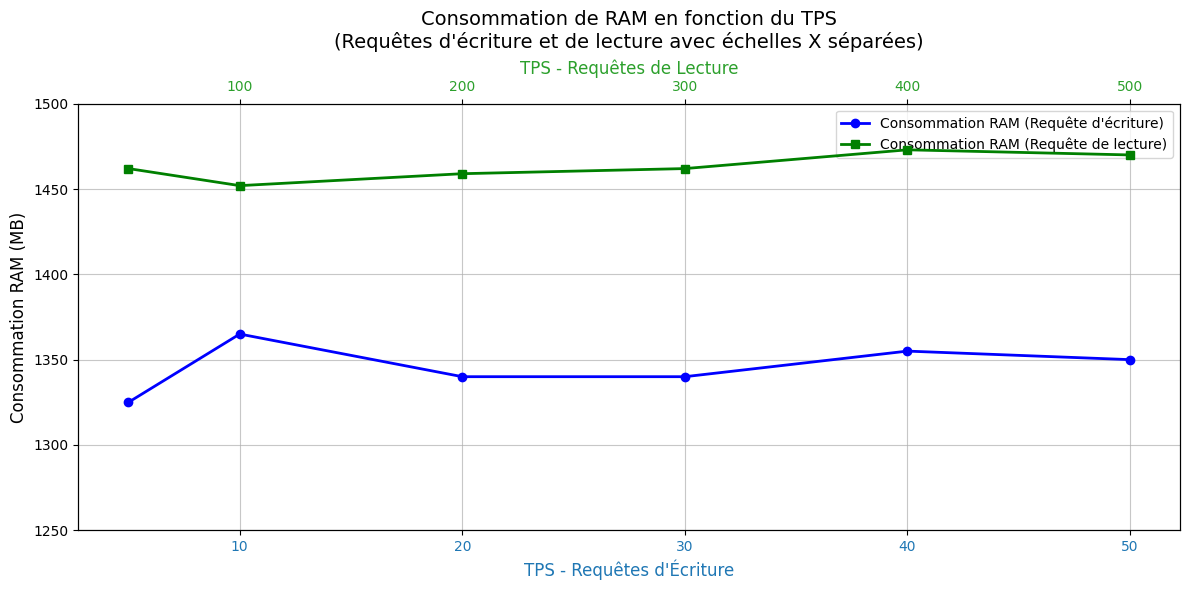}
\caption{CPU consumption based on TPS.}
\label{fig:test6}
\end{figure}

This test examines RAM consumption for 2000 transactions, varying the TPS from 5 to 50. The results (Figure~\ref{fig:test6}) indicate stable consumption: for read requests, it oscillates between 1400.3 MB and 1431.6 MB, and for write requests, between 1332.1 MB and 1361.5 MB.

This constancy suggests that Hyperledger Fabric's memory management is efficient, even under variable loads, an asset for our platform where resource stability is essential. This stability allows for anticipating suitable hardware configurations without oversizing, thus optimizing operating costs. Measurements were carried out on peer and CouchDB containers, with continuous monitoring to capture memory consumption variations.


\section{Discussion and open issues}
\label{sec: 7}
The collaboration of the five protocols of our platform guarantees a reliable and secure system that meets the requirements of the health sector and offers several advantages, such as:
\begin{itemize}
    \item \textbf{Confidentiality :}
    The proposed system guarantees the confidentiality of health data, ensuring that only authorized entities can access it. For this purpose, the EHR-ACCESS protocol does not authorize any actor to consult EHRs, except under the authorization of the patient. Even the latter must authorize himself to consult his EHR for authentication outside the blockchain. 
    \item \textbf{Integrity :}
    This platform gives more visibility and confidence to users who can verify the integrity of any EHR added to the system using the EHR-IV protocol. This functionality is very useful in the field of public health since certain EHRs can be used as tangible proof in the context of litigation or social compensation.
    \item \textbf{Traceability :}
   Traceability is ensured by this platform, which stores information relating to the consultation of each EHR in the system for operational purposes during audit or investigation procedures. 
    \item \textbf{Extensibility :}
    The system, as it is designed, is extensible because:
    \begin{itemize}
        \item Using the EHR-RP, any entity can at any time request to integrate the platform on the condition of proving its identity with the Health Authority to be added to the relevant category among those defined in the system.
        \item By using the EHR-PUB, all the actors of the system can add EHRs. Also, given the decentralized storage mode on which this platform is based, the storage capacity of the system is very large, and it is equal to the sum of the storage capacity of all the health providers that take part in this ecosystem.
    \end{itemize}
    \item \textbf{Authentication :}
    Authentication is essential in a secure data-sharing system in the healthcare sector. Therefore, the design of the proposed system is based on mixed (centralized and decentralized) authentication through the RP protocol; each entity will necessarily go through the Health Authority to create its identity and prove its ability to join a particular category in the system, in a centralized way.\\ Then, at each new interaction with the system, authentication will still be required, but this time it will be done in a decentralized way via the large distributed registry (blockchain), based on the information entered during the registration phase.
    \end{itemize}
    It should be noted that the present system will be able to manage a medical system in its normal daily functions. However, to automate some specific needs of the sector, three other protocols need to be considered, which involve identifying the real needs of the different health providers. These three protocols are enumerated as follows: 
\begin{enumerate}
\item \textbf{Medical emergency :}
Medical emergencies are among the most important services in the medical sector. These services take care of patients in critical or distressed situations.  Therefore, it is paramount to provide a protocol that manages the sharing of information when the patient is unable to do it. This protocol should ensure that health providers concerned with emergencies have access to the necessary information without prior authorization.  
\item \textbf{Investigation :}
Such a protocol will allow specialized and vital entities to investigate a particular fact if it has a relationship with the system. Several situations may require the use of this protocol, such as police investigations and Health Authority investigations for suspicious uses of the platform (usurpation of account, misuse of the medical emergency protocol, etc).
\item \textbf{Research and data exploitation :}
This protocol is responsible for the use of health data in scientific research, statistics on frequent diseases, monitoring of the epidemiological situation, the strategic orientation of the pharmaceutical industry, etc. 
\end{enumerate}


\section{Conclusion}
\label{sec: conc}
By leveraging a private blockchain with smart contracts, we aim to establish a secure and transparent framework that grants patients exclusive control over their EHRs, aligning with privacy regulations like GDPR. In contrast to traditional public key infrastructures,  blockchain technology stood out for its ability to provide a patient-centric model and robust security guarantees. Our proposed architecture presents a practical and robust solution to enhance healthcare data management, enabling secure and efficient sharing of electronic health records (EHRs). By addressing key functional needs, such as data confidentiality, integrity, traceability, and authentication, our platform seeks to strengthen healthcare data sharing and decision-making processes. The significance of secure data sharing in the healthcare sector cannot be overstated. With our architecture, patient data remains confidential, ensuring that only authorized entities can access sensitive medical information. Moreover, the integrity of EHRs is guaranteed, instilling trust and confidence in healthcare professionals and patients alike.

We believe that the impact of our proposed architecture on the healthcare industry can be transformative. Fostering collaboration and data sharing among healthcare providers, it has the potential to revolutionize patient care and streamline medical operations. Moreover, the extensibility of the system ensures its adaptability to future changes and emerging technologies in the healthcare domain.



\section*{Acknowledgment}
The authors would like to thank Mohamed Maamri and Adel Kebout for their work in the implementation of the experimental component of this work.

 \section*{Declaration of Competing Interest} 
 The authors declare that they have no known competing financial interests or personal relationships that could have appeared to influence the work reported in this paper.

\section*{CRediT authorship contribution statement}

\textbf{Tayeb Kenaza}: Conceptualization, Validation, Methodology, Writing – original draft, Supervision.  
\textbf{Islam Debicha}: Validation, Writing – review \& editing, Supervision.  
\textbf{Youcef Fares}: Validation, Writing – review \& editing, Supervision. 
\textbf{Mehdi Sehaki}: Validation, Writing – review \& editing, Supervision. 
\textbf{Sami Messai}: Software, Visualization, Writing – original draft.

\bibliography{MyBib}   

%
%




\end{document}